\documentclass{ptapap}
\usepackage{caption}

\author{\'Aron Juh\'asz}[CAMK,FICT]
\author{L\'aszl\'o Moln\'ar}[CAMK]
\affil[CAMK]{E{\"o}tv{\"o}s University, 
 Budapest, Hungary}
\affil[FICT]{Konkoly Observatory, MTA, CSFK, 
 Budapest, Hungary}

\title{From PanSTARRS Candidates to New RR~Lyraes in the \textit{K2} Mission}

\begin{document}

\maketitle

\begin{abstract}
In this paper we report the discovery of 35 new RR~Lyrae
variables. These stars were found by a special searching technique. We
crossmatched the catalog of the PanSTARRS (PS) sky survey with {\it
  K2} space photometry data to the validate candidates. It turns out
that this technique can find eclipsing binaries as well. 
\end{abstract}

\section{Introduction}

The PS $3\pi$ survey contains more than 25 million
targets. \cite{Hernitschek} published a paper which supplements the
catalog of the survey. In their study, they improved an existing
method \citep{Sesar2010} to add unique RR~Lyrae probabilities to each
target. They tagged nearly $400\,000$ stars above a $1$\,\%
probability of being an RR~Lyrae variable. More than a thousand of
them were measured by the {\it K2} mission of {\it Kepler}. This gives
us the opportunity to search formerly unknown RR~Lyrae variables using all the archive data of the first 12 fields of {\it K2}. 

\section{Targets and New Variables in the \textit{K2} Field}

We divided the potential targets into two categories. We estimated
that 271 stars could be near enough to the main targets to appear on
the target pixel files (TPF) as background stars. From these we were able to identify new RR Lyraes and also
made light curves. We examined 939 main targets from {\it K2} data. These are usually known stars but
some of them were misclassified. Our main goal was to find new
RR~Lyraes in the target images of {\it K2}.  

First, we examined the 271 background targets. We paired the coordinates 
of the PS targets with the {\it K2} TPFs, and then we 
performed photometry of these stars with the PyKE
software \citep{Stillbarclay}. We found 59 RR~Lyraes and 21 among of them were not previously cataloged. The results shown on Table~\ref{tab1}.  

Secondly, based on the directly measured 939 {\it K2} targets, we
found 779 RR~Lyraes (Table~\ref{tab2}). We identified 14 new RR~Lyraes
among these formerly known stars. These targets were misclassified as
mostly eclipsing binaries but the {\it K2} data show clearly RR~Lyrae
shaped light curves.  

\begin{figure}[!h]

\begin{minipage}{.5\textwidth}
 \begin{center}
 \begin{tabular}{ | l | r | r |}
   \hline
 Type			& Number 	& \% 		\\ \hline   
 RRab			& 54		& 19.9\% 	\\ \hline 
 RRc			& 4		& 1.5\%		\\ \hline 
 RRd			& 1		& 0.4\%		\\ \hline
 W UMa			& 38		& 14.0\%	\\ \hline
 $\mathbf{\beta}$ Lyr	& 9		& 3.3\%		\\ \hline
 $\mathbf{\beta}$ Per	& 11	 	& 4.1\%		\\ \hline 
 Unidentified		& 90		& 33.2\%	\\ \hline
 Out of TPF		& 64		& 23.6\%	\\ \hline
 All			& 271		& 100.0\%	\\ \hline
 \end{tabular}
\captionof{table}{Background targets.\label{tab1} }
\end{center} 
\end{minipage}
\begin{minipage}{.5\textwidth}
 \begin{center}
 \begin{tabular}{ | l | r | r |}
 \hline
 Type			& Number 	& \% 		\\ \hline   
 RRab			& 635		& 68.2\% 	\\ \hline 
 RRc			& 123	 	& 13.1\%	\\ \hline 
 RRd			& 21		& 2.6\%		\\ \hline
 $\mathbf{\delta}$ Cep	& 6		& 0.6\%		\\ \hline
 W UMa			& 30		& 3.2\%		\\ \hline
 $\mathbf{\beta}$ Lyr	& 23	 	& 2.5\%		\\ \hline 
 $\mathbf{\beta}$ Per	& 2		& 0.2\%		\\ \hline
 Unidentified		& 99		& 10.5\%	\\ \hline
 All			& 939		& 100.0\%	\\ \hline
 \end{tabular}
 \captionof{table}{Directly measured targets.\label{tab2}}
\end{center} 

\end{minipage}
\end{figure}

\section{Field 0 – the Case of the Anti-Center}

RR~Lyraes appear in the Galactic disk but mostly in the direction of
the Bulge. Nonetheless, the PS data contain numerous candidates in the direction of the Galactic anti-center. The first observing zone of the {\it K2} mission (Field 0) partially covered this area. The Galactic
plane contains a lot of binaries at all Galactic longitudes. The
majority of {\it K2} observed PS sources in Field 0 also turned out to
be eclipsing binaries, suggesting that the large density of disk
RR~Lyrae candidates in the PS catalog may come from confusion between
these two sources. The second iteration of the catalog
\citep{Sesar2017} excluded the Galactic plane with high extinction,
and thus removed the most problematic areas of the survey.

\section{Summary}

Using the {\it K2} archive data, we revealed the true nature of 14
previously misclassified RR~Lyraes and found 21 brand new ones (19
RRab, 2 RRc). Furthermore, we identified numerous new eclipsing
binaries. We found that among the PS false positive targets, eclipsing
binaries appear frequently. Based on {\it K2} Field 0 the Galactic
disk could be particularly overrepresented with these impostors. It
suggests that the PS catalog also provides the opportunity to find
binaries in a similar way. 

\acknowledgements{The research have been supported by the the
  \'UNKP-17-3 program of the Ministry of Human Capacities of Hungary,
  and the Hungarian National Research, Development and Innovation
  Office (NKFIH) grants K-115709, PD-116175, and the Lend{\"u}let
  LP2014-17 grant of the Hungarian Academy of Sciences. LM was
  supported by the J\'anos Bolyai Research Scholarship of the
  Hungarian Academy of Sciences.} 

\bibliographystyle{ptapap}
\bibliography{ptapapdoc}

\end{document}